\documentclass[twocolumn,floatfix,superscriptaddress,a4paper,
               showpacs,showkeys,nofootinbib,reprint,prl]{revtex4}
\usepackage{epsfig}
\usepackage{latexsym}
\usepackage{xspace}
\usepackage{hyperref}
\usepackage[latin2]{inputenc}
\usepackage{indentfirst}
\usepackage{enumerate}
\usepackage{color}

\usepackage{amsmath}
\usepackage{amssymb}
\usepackage[english]{babel}
\usepackage{url}
\newcommand{\bs}{\boldsymbol}

\begin{document}

%TC:ignore

\title{
van der Waals Interactions in Hadron Resonance Gas: \\
From Nuclear Matter to Lattice QCD
}

\author{Volodymyr Vovchenko}
\affiliation{
Frankfurt Institute for Advanced Studies, Goethe Universit\"at Frankfurt,
D-60438 Frankfurt am Main, Germany}
\affiliation{
Institut f\"ur Theoretische Physik,
Goethe Universit\"at Frankfurt, D-60438 Frankfurt am Main, Germany}
\affiliation{
Department of Physics, Taras Shevchenko National University of Kiev, 03022 Kiev, Ukraine}

\author{Mark I. Gorenstein}
\affiliation{
Bogolyubov Institute for Theoretical Physics, 03680 Kiev, Ukraine}
\affiliation{
Frankfurt Institute for Advanced Studies, Goethe Universit\"at Frankfurt,
D-60438 Frankfurt am Main, Germany}

\author{Horst Stoecker}
\affiliation{
Frankfurt Institute for Advanced Studies, Goethe Universit\"at Frankfurt,
D-60438 Frankfurt am Main, Germany}
\affiliation{
Institut f\"ur Theoretische Physik,
Goethe Universit\"at Frankfurt, D-60438 Frankfurt am Main, Germany}
\affiliation{
GSI Helmholtzzentrum f\"ur Schwerionenforschung GmbH, D-64291 Darmstadt, Germany}

\begin{abstract}
An extension of the ideal 
hadron resonance gas (HRG) model
is constructed
which includes
the attractive and repulsive van der Waals (VDW) interactions between baryons.
This VDW-HRG model yields the nuclear liquid-gas transition at low temperatures and high baryon densities.
The VDW parameters $a$ and $b$
are fixed by the ground state properties of nuclear matter,
and the temperature dependence of
various thermodynamic observables at zero chemical potential
are calculated
within VDW-HRG model.
Compared to the ideal 
HRG model, the inclusion of VDW interactions between baryons leads to a qualitatively different behavior of second and higher moments of fluctuations of conserved charges, in particular in the so-called crossover region
$T\sim 140 - 190$~MeV.
For many observables this behavior resembles closely the results obtained from lattice QCD simulations.
This hadronic model also predicts nontrivial behavior of net-baryon fluctuations in the region of phase diagram probed by heavy-ion collision experiments.
These results imply that VDW interactions play a crucial role in thermodynamics of hadron gas. 
Thus, the commonly performed
comparisons of the ideal HRG model with the lattice 
and heavy-ion
data may lead to misconceptions and misleading conclusions.
%, and should therefore be treated with extreme care.
\end{abstract}



\maketitle

%TC:endignore 

The thermodynamic properties of strongly interacting matter at zero chemical potential and finite temperature have been computed using Monte Carlo simulations in lattice QCD~\cite{Borsanyi:2013bia,Bazavov:2014pvz}.
A crossover is observed~\cite{Aoki:2006we}
in the temperature range of $140-190$~MeV.
At lower temperatures, $T \sim 100-150$~MeV, QCD exhibits features 
similar to simple ideal 
hadron resonance gas (IHRG)
which successfully reproduces many lattice observables~\cite{Borsanyi:2011sw,Bazavov:2012jq,Bellwied:2015lba,Bellwied:2013cta}.
In the crossover region, however, the agreement between IHRG and lattice QCD deteriorates. The breakdown of the IHRG model especially concerns the higher order fluctuations and correlations of conserved charges~\cite{Bazavov:2013dta},
resulting in statements 
%that this behavior proves 
that hadrons melt quickly %with increasing temperature 
and are basically absent at $T>160$~MeV~\cite{Karsch:2013naa}.
In this Letter, it is shown that these conclusions are inconclusive.
van der Waals (VDW) interactions between baryons play a crucial role for the thermodynamics of hadron fluid at sufficiently high temperatures.
As a result,
the qualitative features of the thermodynamics of interacting HRG 
appear to be close to lattice results in the crossover region.
The results also have important phenomenological relevance for 
heavy-ion collision experiments where measurements of conserved charges fluctuations have been suggested as probes for chemical freeze-out~\cite{Bazavov:2012vg,Borsanyi:2014ewa} or the QCD critical point~\cite{Stephanov:1998dy}.

The IHRG model
does not capture the VDW nature of nucleon-nucleon interaction and, thus, fails
to describe the properties of nuclear matter at small temperatures and large baryon densities.
This shortcoming of this common HRG model is usually considered to be of minor significance when applied to ultrarelativistic heavy-ion collisions or to lattice data,
although recently possible relevance for
fluctuations was pointed out~\cite{Fukushima:2014lfa}.
The repulsive part of VDW interactions had often been included into HRG by means of an excluded-volume (EV) procedure 
%introduced in 
\cite{Rischke:1991ke}, usually assuming identical EV interactions between all hadron pairs~\cite{BraunMunzinger:1999qy}.
%Such a modification, however, is still insufficient to describe basic nuclear matter properties.
%This model is, however, not able to describe the nuclear liquid-gas phase transition in nuclear matter.
The grand canonical ensemble formulation of the full VDW equation with both attractive and repulsive interactions, and including quantum statistics, was developed in
Refs.~\cite{Vovchenko:2015xja,Vovchenko:2015vxa,Vovchenko:2015pya} for single-component systems.
In these works, the basic features of nuclear matter have been successfully described by the VDW equation with Fermi statistics for nucleons.
The VDW parameters $a$ and $b$ were uniquely fixed by reproducing the saturation density $n_0 = 0.16$~fm$^{-3}$ and binding energy $E/A = -16$~MeV of the ground state of nuclear matter.
For nucleons the values $a = 329$~MeV$~$fm$^3$ and $b = 3.42$~fm$^3$~were obtained in Ref.~\cite{Vovchenko:2015pya}, and also later in Ref.~\cite{Redlich:2016dpb}.
The resulting model predicts a liquid-gas first-order phase transition in symmetric nuclear matter with a critical point located at $T_c \simeq 19.7$~MeV and $\mu_c \simeq 908$~MeV ($n_c \simeq 0.07$~fm$^3 = 0.45\,n_0$).

In the following a minimal extension of IHRG model, which includes the VDW interactions between (anti)baryons, is described. We refer to this model as VDW-HRG and it is based on the following assumptions:
%\begin{enumerate}

1. VDW interactions are assumed to exist between all pairs of baryons and between all pairs of antibaryons. The VDW parameters $a$ and $b$ for all (anti)baryons are assumed to be equal to those of nucleons, as obtained from the fit to the ground state of nuclear matter.

2. The baryon-antibaryon, meson-meson, and meson-(anti)baryon VDW interactions are neglected.
%\end{enumerate}

In a sense the present VDW-HRG model is a ``minimal-interaction'' extension of the IHRG model, which describes the basic properties of nuclear matter.
%{\bf and introduces minimum amount of new parameters.}
Whether significant VDW interactions exist between hadron pairs other than (anti)baryons is not clearly established.
For instance, it has been argued that short-range interactions between baryons and antibaryons may be dominated by annihilation processes and not by repulsion~\cite{Andronic:2012ut},
and this is our motivation to exclude VDW terms for them in this study.
The presence of significant mesonic eigenvolumes, comparable to those of baryons, leads to significant suppression
of thermodynamic functions in the crossover region at $\mu_B=0$, 
which is at odds with lattice data (see Refs.~\cite{Andronic:2012ut,Vovchenko:2014pka}).
The attractive interactions involving mesons, on the other hand,
normally lead to resonance formation~\cite{Venugopalan:1992hy},
which are already included in HRG by construction.
For these reasons we neglect the meson-related VDW interactions in this study.
%It is clear from the listed assumptions that 
The VDW-HRG consists of three
sub-systems: Non-interacting mesons, VDW baryons, and VDW antibaryons. The total pressure reads
\begin{equation}
p(T,\bs \mu) = P_M(T,\bs \mu) + P_B(T,\bs \mu) + P_{\bar{B}}(T,\bs \mu),
\end{equation}
with
\begin{align}
P_M(T,\bs \mu) & =
\sum_{j \in M} p_{j}^{\rm id} (T, \mu_j) \\
\label{eq:PB}
P_B(T,\bs \mu) & =
\sum_{j \in B} p_{j}^{\rm id} (T, \mu_j^{B*}) - a\,n_B^2 \\
\label{eq:PBBar}
P_{\bar{B}}(T,\bs \mu) & =
\sum_{j \in \bar{B}} p_{j}^{\rm id} (T, \mu_j^{\bar{B}*}) - a\,n_{\bar{B}}^2,
\end{align}
where $M$ stands for mesons, $B$ for baryons, and $\bar{B}$ for antibaryons, $p_{j}^{\rm id}$ is the Fermi or Bose ideal gas pressure,
$\bs \mu=(\mu_B,\mu_S,\mu_Q)$ are the chemical potentials which regulate the average values of net baryon number $B$, strangeness $S$, electric charge $Q$,
$\mu_j^{B(\bar{B})*} = \mu_j - b\,P_{B(\bar{B})} - a\,b\,n_{B(\bar{B})}^2 + 2\,a\,n_{B(\bar{B})}$,
and $n_B$ and $n_{\bar{B}}$ are, respectively, total densities of baryons and antibaryons.

The calculation of mesonic pressure $P_M(T,\bs \mu)$ is straightforward.
%For baryons and antibaryons the situation is more complicated: 
The shifted chemical potentials $\mu_j^{B(\bar{B})*}$ of (anti)baryons depend explicitly on (anti)baryon pressure $P_{B(\bar{B})}$ and on total (anti)baryon density $n_{B(\bar{B})}$. By taking the derivatives of $P_{B(\bar{B})}$ with respect to the baryochemical potential one obtains additional equations for $n_{B(\bar{B})}(T,\bs \mu)$
\begin{align}
\label{eq:nB}
n_{B(\bar{B})}
%(T,\bs \mu) 
~ =~
(1 - b \, n_{B(\bar{B})})  \sum_{j \in B(\bar{B})} n_{j}^{\rm id} (T, \mu_j^{B(\bar{B})*})~.
\end{align}

At given $T$ and $\bs \mu$, Eqs.~\eqref{eq:PB}-\eqref{eq:nB}
are solved numerically, giving $P_{B(\bar{B})}(T,\bs \mu)$ and $n_{B(\bar{B})}(T,\bs \mu)$.
% are obtained.
%Finally, the 
Entropy density is calculated as $s = (\partial p / \partial T)_{\mu}$, and energy density is obtained from Gibbs relation.

\begin{figure*}[t]
\begin{center}
\includegraphics[width=0.49\textwidth]{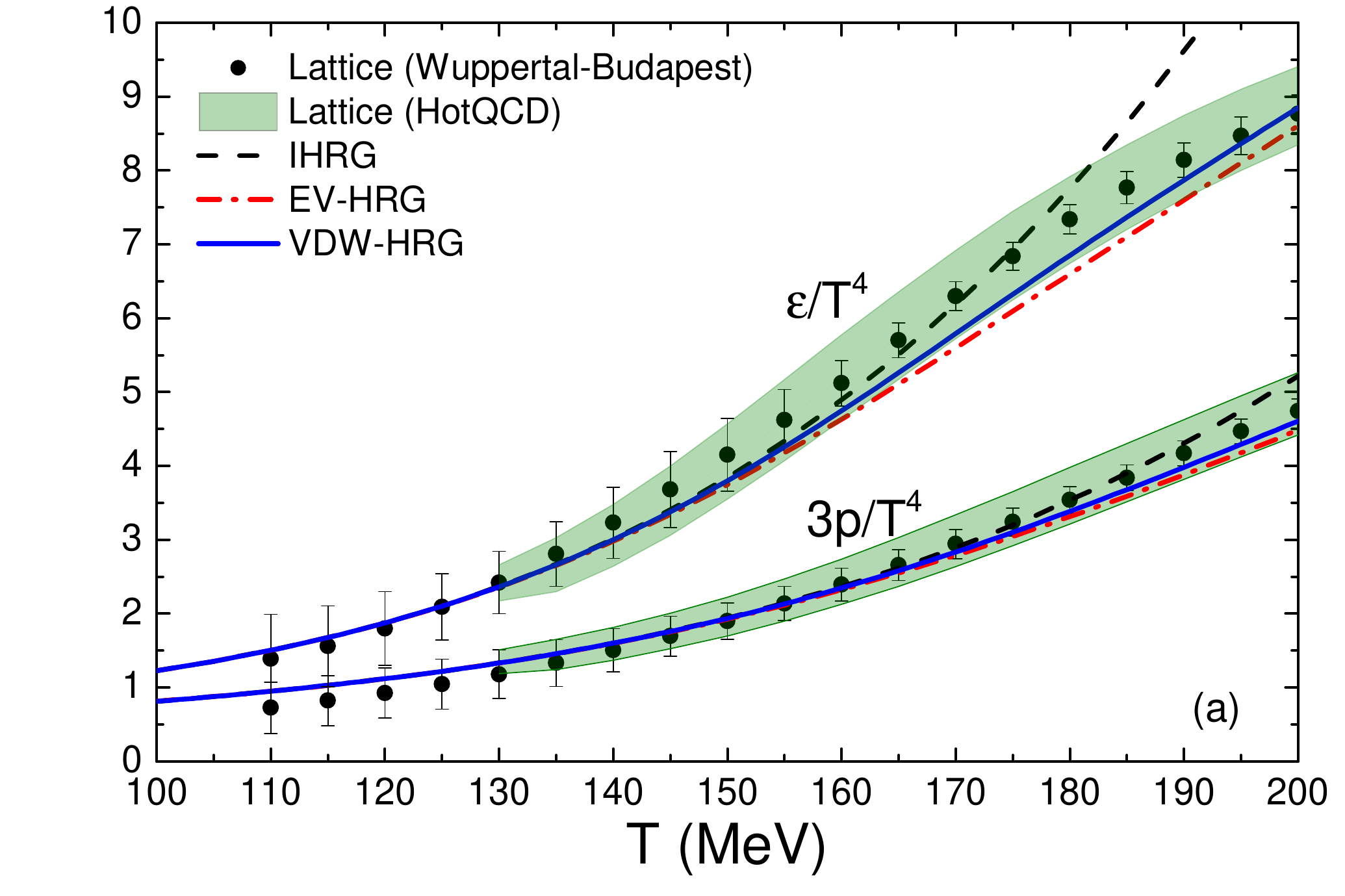}
\includegraphics[width=0.49\textwidth]{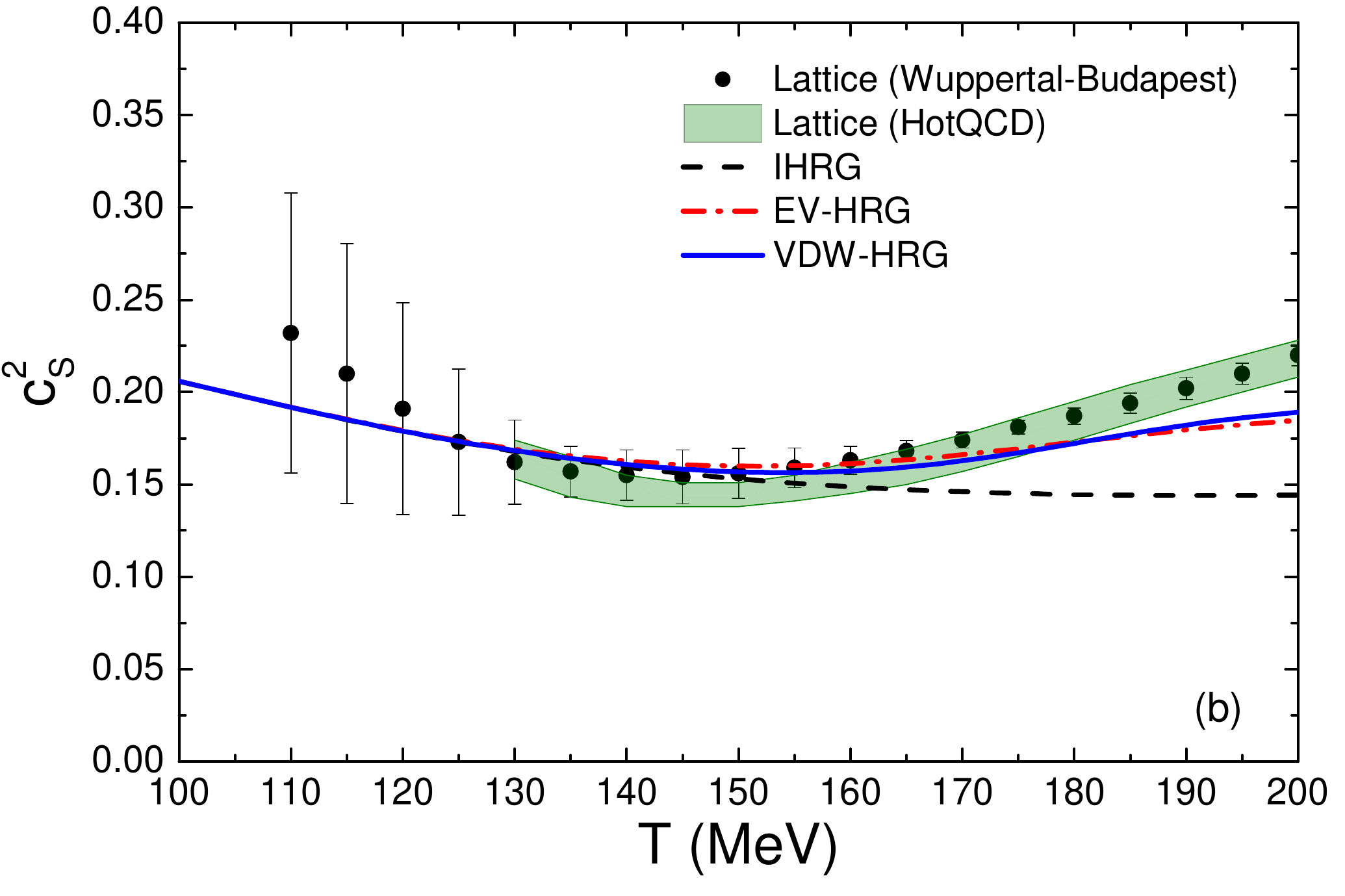}
\caption{The temperature dependence of
(a) scaled pressure and energy density,
and of
(b) the square of the speed of sound
at zero chemical potential, as calculated within IHRG (dashed black lines), EV-HRG with $b = 3.42$~fm$^3$ (dash-dotted red lines), and VDW-HRG with $a = 329$~MeV~fm$^3$ and $b = 3.42$~fm$^3$ (solid blue lines). Lattice QCD results of Wuppertal-Budapest~\cite{Borsanyi:2013bia} and HotQCD~\cite{Bazavov:2014pvz} collaborations are shown, respectively, by symbols and green bands.
}
\label{fig:thermod}
\end{center}
\end{figure*}

The present calculations include all established strange and non-strange hadrons which are listed in the Particle Data Tables~\cite{Agashe:2014kda}, %and have a confirmed status there. 
with exception of $\sigma$ and $\kappa$ mesons~\cite{Broniowski:2015oha,Friman:2015zua}.
%The exceptions are $\sigma$ and $\kappa$ mesons, following the cancellation arguments~\cite{Broniowski:2015oha,Friman:2015zua}.
The finite widths of the resonances are included by means of 
an additional mass integration over their %corresponding 
Breit-Wigner shapes. 
We employ the HRG code used in~\cite{Vovchenko:2015cbk,Vovchenko:2015idt}, modified to include the VDW interactions between (anti)baryons.
The temperature dependence of the scaled pressure $p/T^4$, energy density $\varepsilon/T^4$, and the speed of sound squared $c_s^2 = d p /d \varepsilon$ calculated at $\bs \mu=0$ within IHRG and VDW-HRG models is compared to the lattice data in Fig.~\ref{fig:thermod}.
%In order 
To clarify the role of attractive and repulsive interactions we also show the calculations,
denoted as EV-HRG,
where the VDW attraction 
%between baryons 
was ``switched off'', i.e. $a=0$. 
%These calculations are denoted as EV-HRG.
%The important difference in our approach compared to earlier studies involving interacting hadrons~\cite{Andronic:2012ut,Vovchenko:2014pka} is that we only include the EV interactions between (anti)baryons, but not between other hadrons.
Since the matter is meson dominated at $\mu_B=0$, and the mesons are modeled as non-interacting, no significant suppression of thermodynamic functions is seen, in contrast to earlier studies~\cite{Andronic:2012ut,Vovchenko:2014pka}, where constant EV interactions between all hadrons were assumed. 
The energy density is somewhat below the lattice data at $T>160$~MeV for VDW-HRG.
This may be 
explained by
missing heavy Hagedorn states which add a significant contribution to the energy density~\cite{Vovchenko:2014pka}.
The temperature dependence of the speed of sound squared, $c_s^2$, is consistent with lattice data
and shows a minimum at $T \sim 155-160$~MeV, in contrast to the IHRG where $c_s^2$ decreases slowly and monotonically. 

\begin{figure*}[t]
\begin{center}
\includegraphics[width=0.49\textwidth]{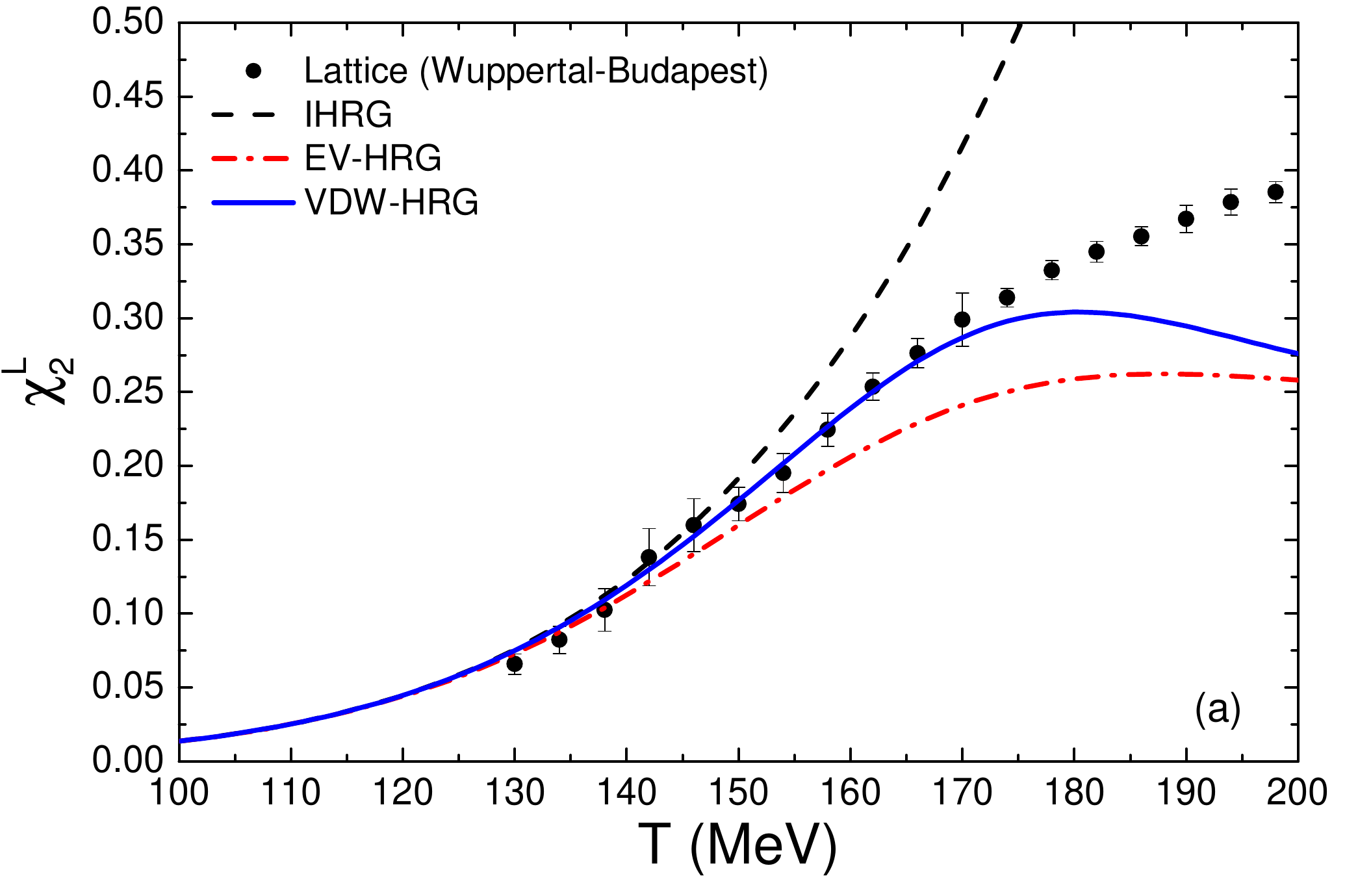}
\includegraphics[width=0.49\textwidth]{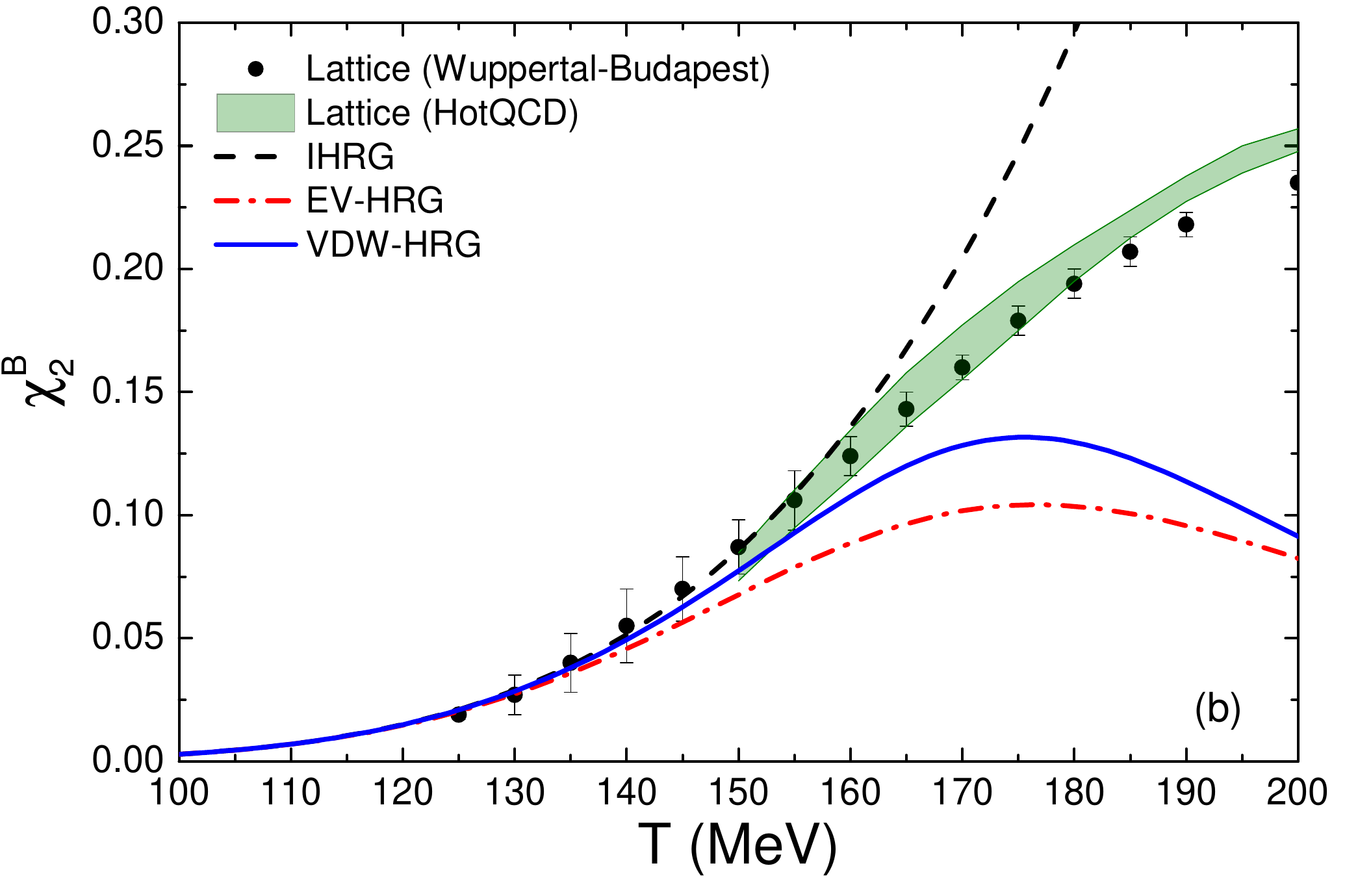}
\includegraphics[width=0.49\textwidth]{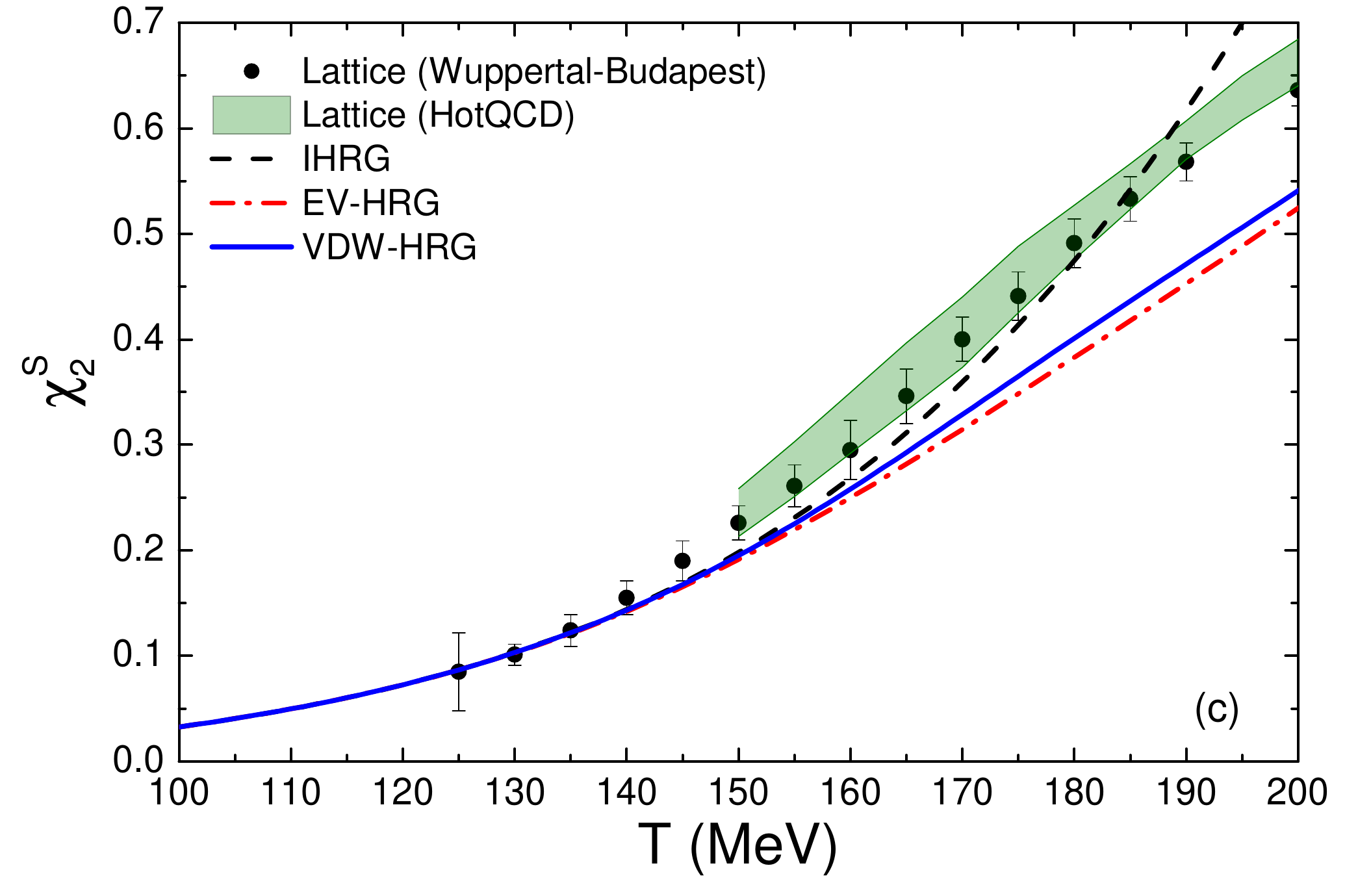}
\includegraphics[width=0.49\textwidth]{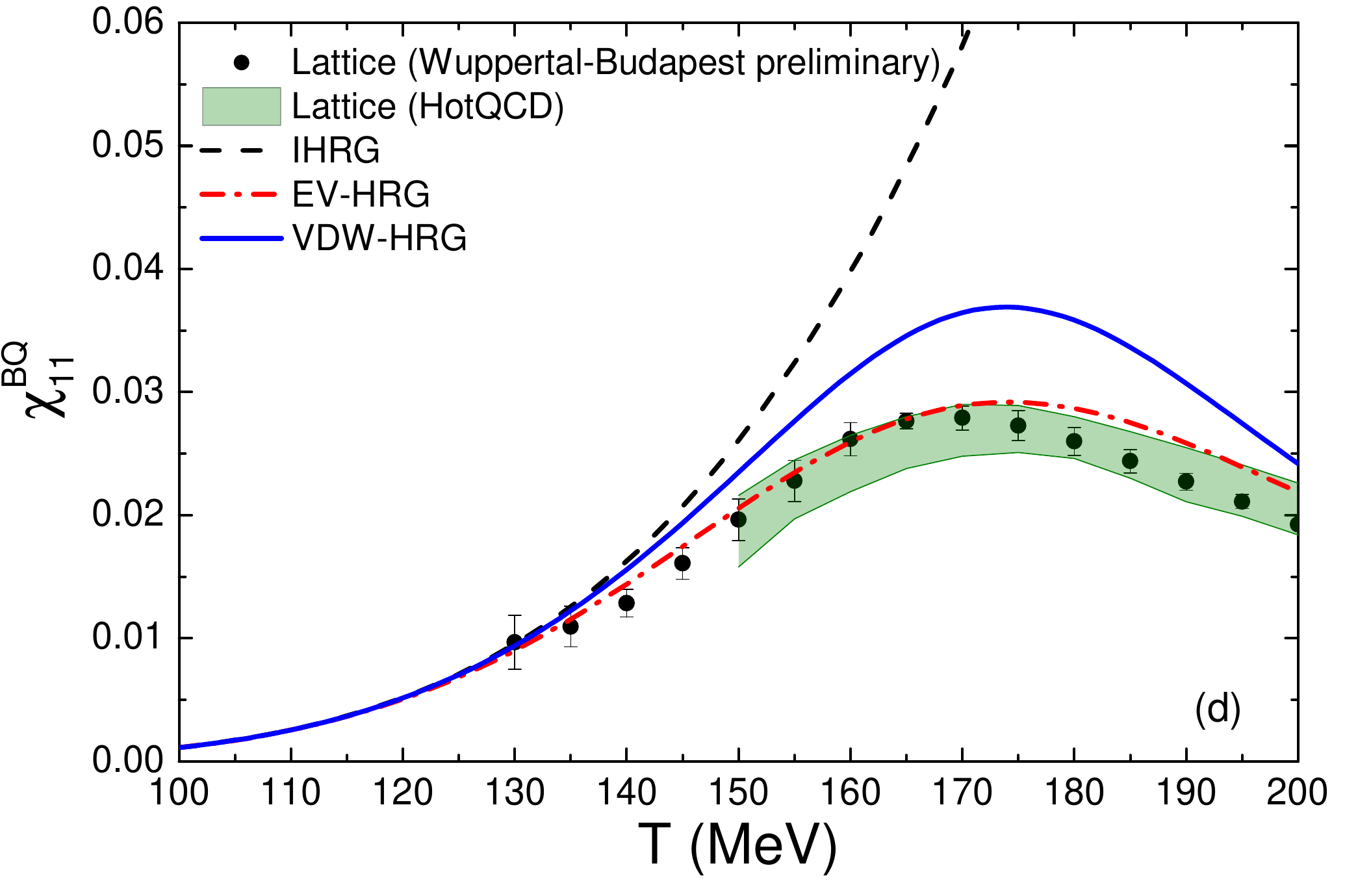}
\caption{The temperature dependence of the 2nd order susceptibilities of conserved charges.
These include (a) net number of light quarks $\chi_2^L$,
(b) net baryon number $\chi_2^B$,
(c) net strangeness $\chi_2^S$,
and (d) baryon-electric charge correlator $\chi_{11}^{BQ}$.
Calculations are done within IHRG (dashed black lines), EV-HRG (dash-dotted red lines), and VDW-HRG (solid blue lines). Lattice QCD results of the Wuppertal-Budapest~\cite{Borsanyi:2011sw,Bellwied:2013cta}~(for $\chi_{11}^{BQ}$ preliminary results~\cite{Borsanyi:2015axp,WBchi11BQ} are used) and HotQCD~\cite{Bazavov:2012jq} collaborations are shown, respectively, by symbols and green bands.
}
\label{fig:chi2}
\end{center}
\end{figure*}

In addition to the thermodynamical functions, the VDW-HRG model allows us to calculate the fluctuations
of conserved charges:
\begin{equation}\label{BSQ}
\chi_{lmn}^{BSQ}~=~\frac{\partial^{l+m+n}p/T^4}{\partial(\mu_B/T)^l \,\partial(\mu_S/T)^m \,\partial(\mu_Q/T)^n}~\,.
\end{equation}
The fluctuations of the net number of light quarks $L = (u+d)/2 = (3B+S)/2$ are also considered.

The temperature dependencies of the 2nd order susceptibilities are shown in Fig.~\ref{fig:chi2}.
These include (a) net number of light quarks $\chi_2^L$,
(b) net baryon number $\chi_2^B$,
(c) net strangeness $\chi_2^S$,
and (d) baryon-electric charge correlator $\chi_{11}^{BQ}$.
The $\chi_2^L$ calculated within the VDW-HRG model shows a very different behavior compared to IHRG at $T>160$~MeV, and agrees well with the
lattice data%from Ref.
~\cite{Borsanyi:2011sw} up to $T=180$~MeV.
A qualitatively similar picture is obtained for $\chi_2^B$.
The qualitative difference between IHRG and VDW-HRG models appears to be driven by the EV interaction terms between (anti)baryons,
while the inclusion of VDW attraction leads to an improved agreement with the lattice data.
%This is illustrated by the EV-HRG calculations, shown by dash-dotted red lines in Fig.~\ref{fig:chi2}.
%
The strangeness susceptibility $\chi_2^S$
is described fairly well by the IHRG model, but
appears to be underestimated by the
VDW-HRG model.
We have also found that the baryon-strangeness correlator 
(not shown in plots) is rather notably underestimated by all considered HRG models.
Does this reflect the presence of hitherto undiscovered strange hadrons?
%, which are presently not included in the particle data tables?
%The inclusion of such states in form of a Hagedorn spectrum~\cite{Lo:2015cca}, or by employing a Quark Model~\cite{Bazavov:2014xya} were shown to improve the agreement between lattice data and IHRG. 
The inclusion of such states was shown to improve the agreement between lattice data and IHRG~\cite{Lo:2015cca,Bazavov:2014xya}. 
%Similar effect can be expected in the VDW-HRG case.
The correlator $\chi_{11}^{BQ}$ between net baryon number and net electric charge has a very different temperature dependence in IHRG and VDW-HRG.
In the IHRG model the $\chi_{11}^{BQ}$ increases rapidly at $T>150$~MeV, in
stark contrast to the lattice data.
In the VDW-HRG model this correlator
has a broad bump with a maximum at $T \sim 160-190$~MeV, showing
a behavior which is in qualitative agreement
to the correlator obtained on
the lattice.

\begin{figure*}[t]
\begin{center}
\includegraphics[width=0.49\textwidth]{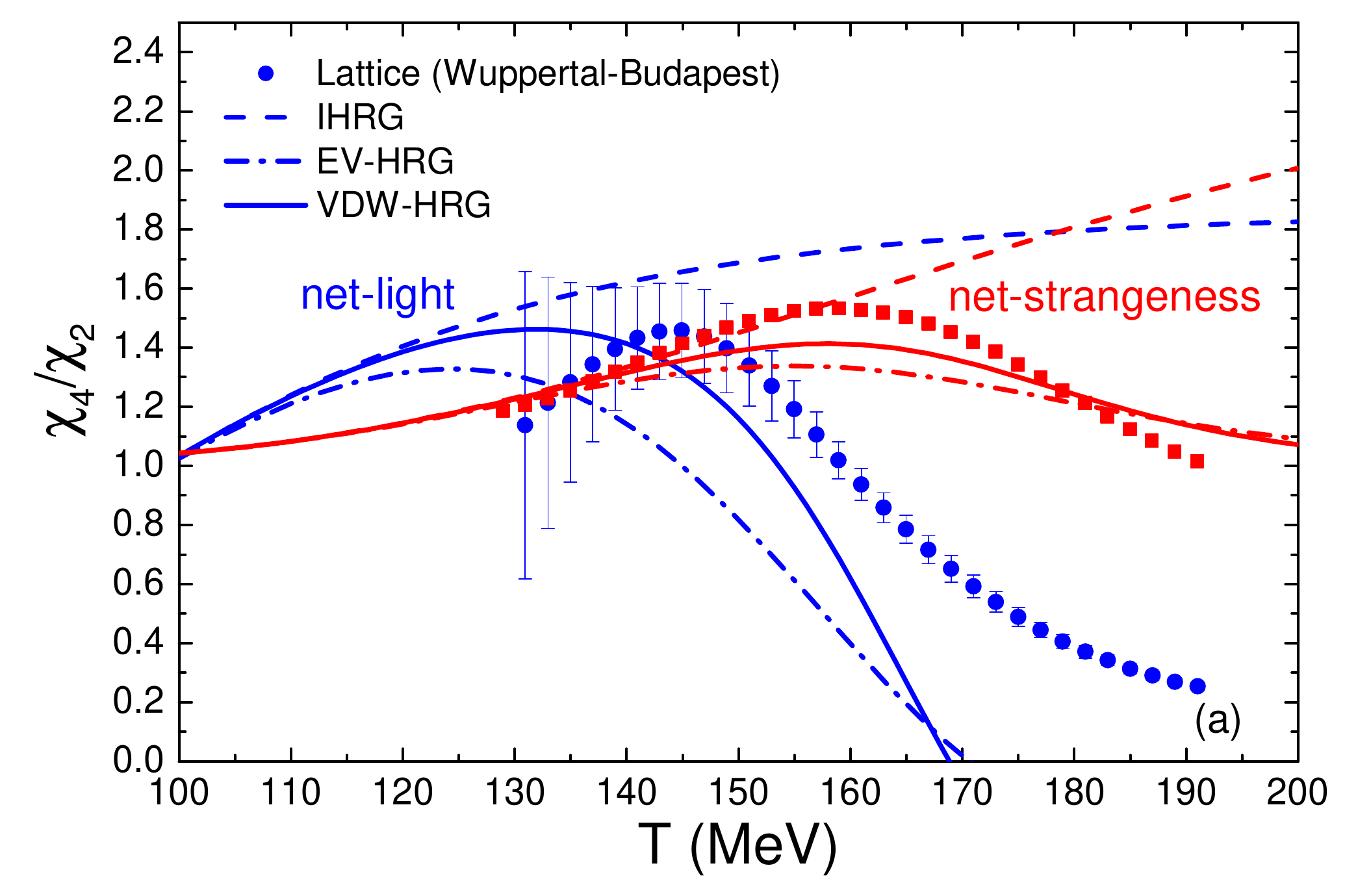}
\includegraphics[width=0.49\textwidth]{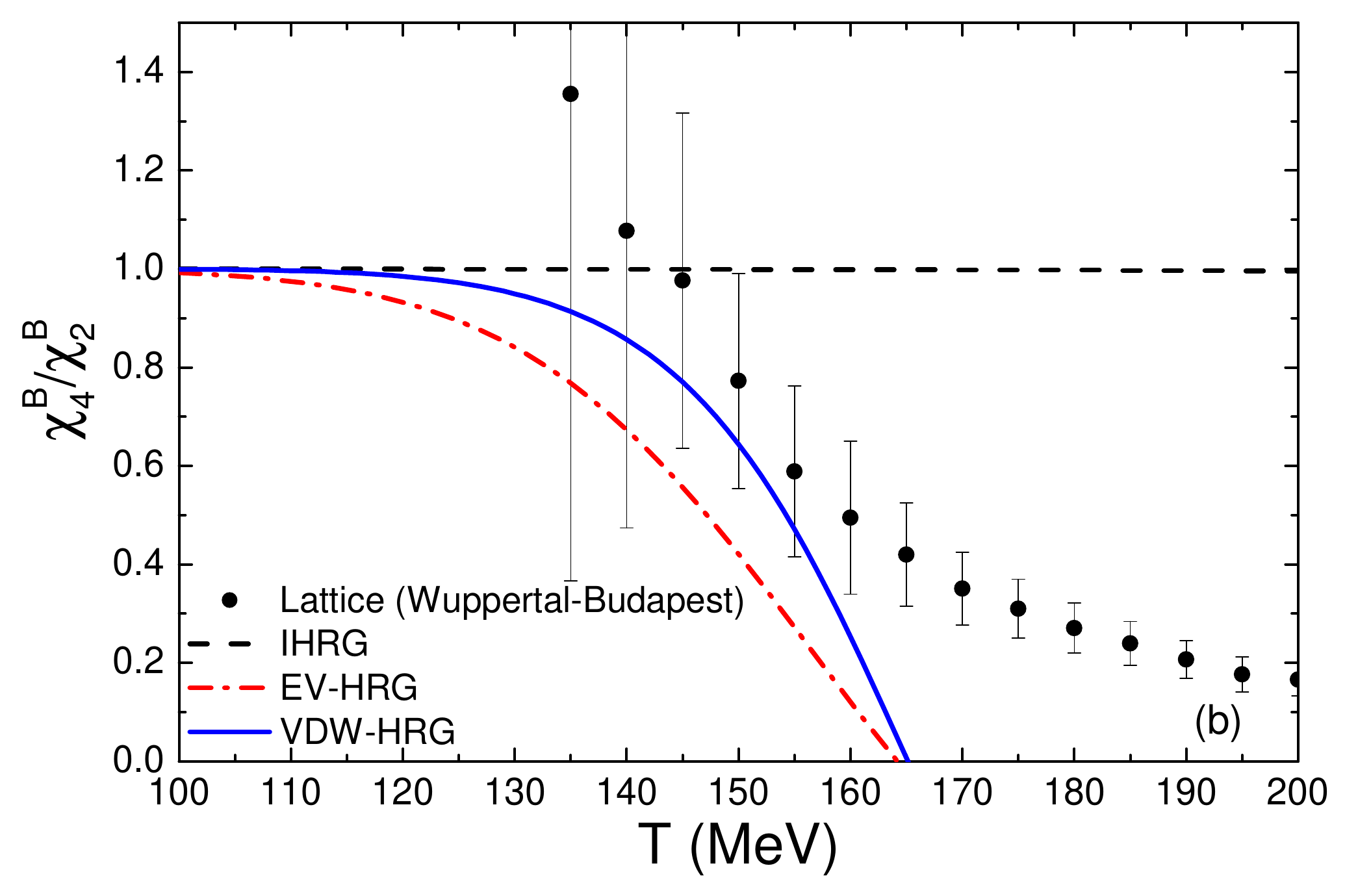}
\includegraphics[width=0.49\textwidth]{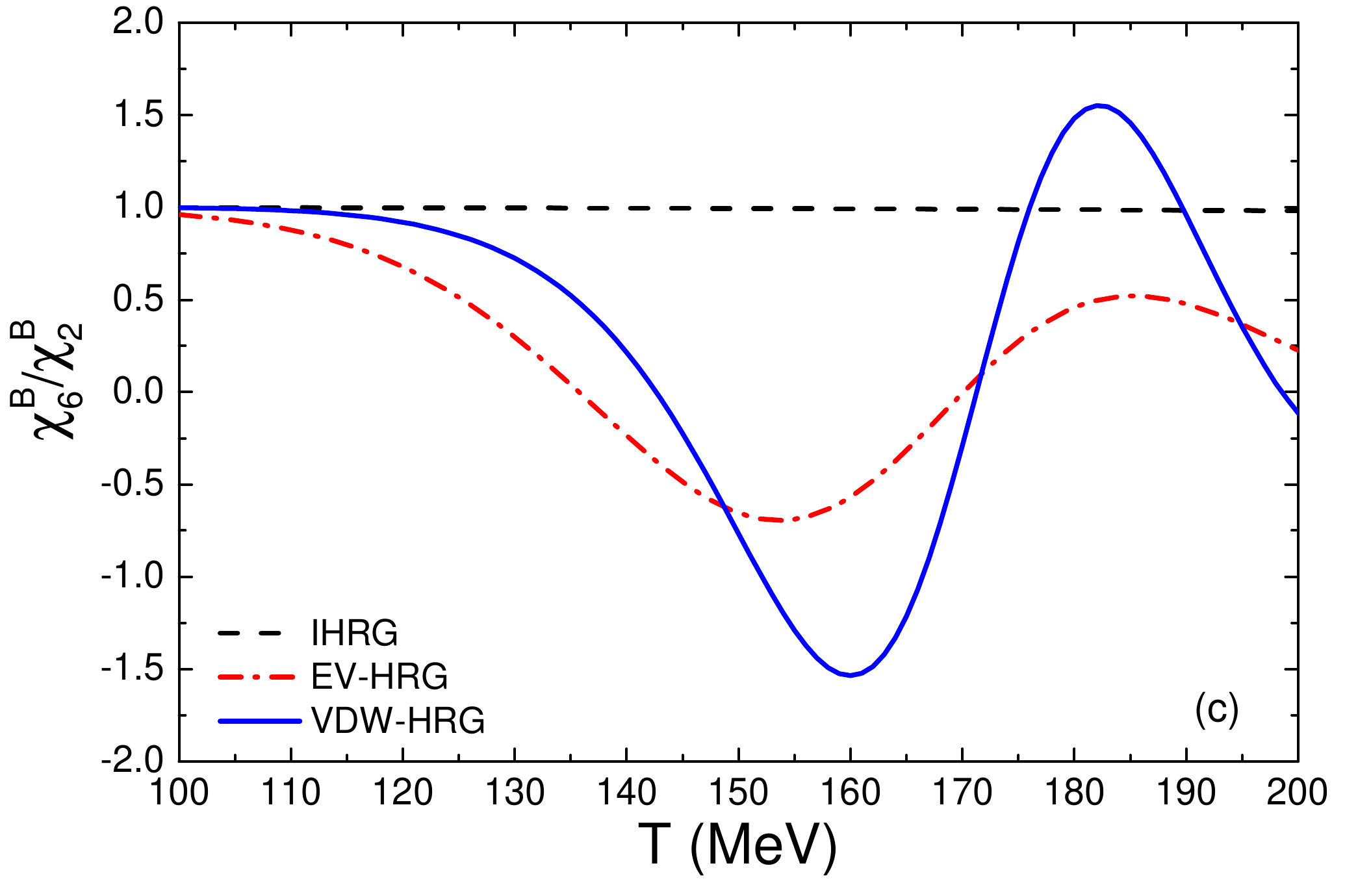}
\includegraphics[width=0.49\textwidth]{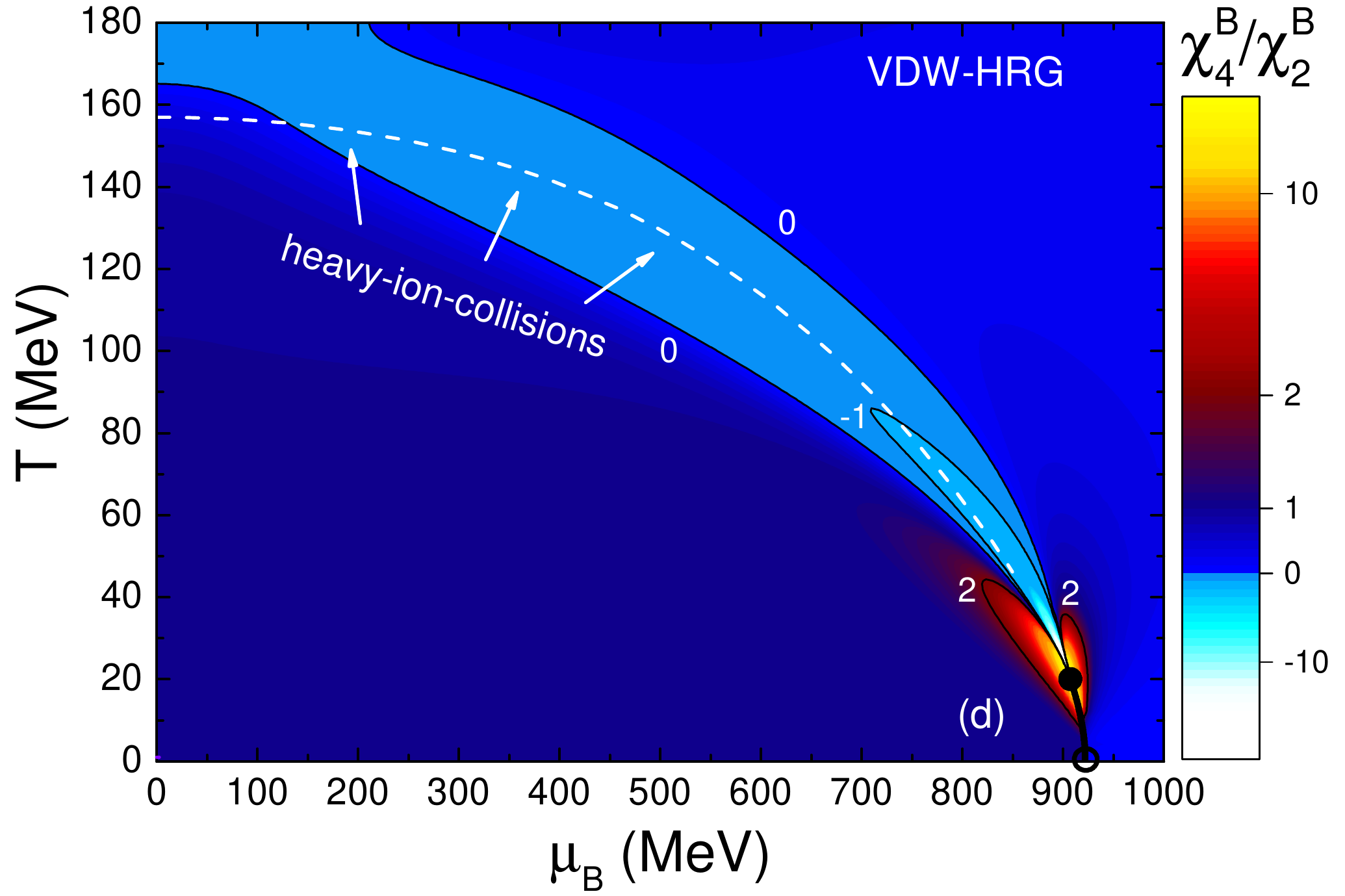}
\caption{The temperature dependence of the $\chi_4/\chi_2$ cumulant ratio for
(a) net number of light/strange quarks,
(b) net baryon number,
and the (c) $\chi_6/\chi_2$ for net baryon number.
Calculations are done within IHRG (dashed lines), EV-HRG (dash-dotted lines), and VDW-HRG (solid lines). The lattice QCD results of the Wuppertal-Budapest collaboration~\cite{Bellwied:2013cta,Bellwied:2015lba} are shown by symbols. (d) The $\chi_4^B/\chi_2^B$ ratio in VDW-HRG in the $T$-$\mu_B$ plane. The possible chemical freeze-out line in heavy-ion collisions taken from~\cite{Vovchenko:2015idt} is sketched by the dashed line.
}
\label{fig:chi4}
\end{center}
\end{figure*}

The higher order fluctuations are also analyzed
and
exhibited in Fig.~\ref{fig:chi4}.
All considered observables show very different behavior between IHRG and VDW-HRG.
The net-light number $\chi_4^L/\chi_2^L$ monotonically increases in the IHRG model and overshoots the lattice data at $T \sim 140$~MeV. The VDW-HRG model, in contrast, yields a non-monotonic behavior with a wide peak at $T \sim 120-145$~MeV,
resembling the lattice data~\cite{Bellwied:2013cta},
which peaks at slightly higher temperature.
%The bump in the lattice data seems to appear at slightly higher temperatures. 
The peak in the $T$-dependence of net strangeness $\chi_4^S/\chi_2^S$ is relatively well reproduced within the VDW-HRG model. In contrast, the IHRG model shows no maximum at all.
It is remarkable that our model shows flavor hierarchy:
The peak for net-light number $\chi_4/\chi_2$ is at smaller temperatures as compared to the peak in net-strangeness. The same result is seen in the lattice data!
It was argued that this observation is related to the flavor separation in the deconfinement transition in QCD~\cite{Bellwied:2013cta}.
Since the VDW-HRG model has only hadronic degrees of freedom,
the present results cast doubt on this interpretation to trace back to deconfinement the observed flavor dependence in $\chi_4/\chi_2$, as well as the presence of the peaks themselves.
Turning to higher-order fluctuations of net-baryon number:
The net-baryon kurtosis, $\chi_4^B/\chi_2^B$, shows the expected Skellam behavior for IHRG model with values very close to unity.
The VDW-HRG model, on the other hand, shows a stark decrease at $T = 130-165$~MeV, i.e.
in the so-called ``crossover region'',
even though the VDW-HRG model does not contain any transition to the quark-gluon degrees of freedom.
The $\chi_4^B/\chi_2^B$ even turns negative at $T>165$~MeV. This decrease of $\chi_4^B/\chi_2^B$ is also seen on the lattice~\cite{Bellwied:2015lba}, although
it starts at higher $T=145$~MeV and $\chi_4^B / \chi_2^B$
does not become negative.
Also, the temperature dependence of the sixth order cumulant ratio $\chi_6^B/\chi_2^B$ is predicted:
The VDW-HRG model exhibits very strong variations and
non-monotonous behavior in the ``crossover region''.
Will a similarly dramatic $T$-dependent behavior be observed in corresponding lattice simulations?

Finally, the kurtosis of net-baryon fluctuations at finite baryon density is explored (Fig.~\ref{fig:chi4}d). For simplicity it is assumed that $\mu_S = \mu_Q = 0$. 
The region of negative $\chi_4^B/\chi_2^B$ at small $\mu_B$ is smoothly connected to the region of the liquid-vapor phase transition in nuclear matter, 
and seems relevant for ``chemical freeze-out'' in heavy-ion collisions (see dashed line). The VDW-HRG model suggests non-monotonic behavior of $\chi_4/\chi_2$ with respect to collision energy, in stark contrast to IHRG~\cite{Alba:2014eba}.
This implies that non-trivial fluctuations of net-baryon number 
in heavy-ion collisions~\cite{Aggarwal:2010wy,Adamczyk:2013dal,Luo:2015ewa} may simply be a manifestation of the nuclear liquid-gas phase transition (see also \cite{Fukushima:2014lfa} and \cite{Mukherjee:2016nhb}). 
Since VDW interactions also affect the thermal fits~\cite{Vovchenko:2015cbk,Satarov:2016peb}, this question demands further studies.

For many observables, the \emph{quantitative} agreement of the VDW-HRG model calculations presented here with the lattice data in the ``crossover region'' is not perfect. This is hardly surprising. Indeed, we have modeled the VDW interactions between baryons in the simplest way possible:
It is assumed that the VDW interactions between all baryons are the same as those between nucleons, as obtained from nuclear matter properties at $T=0$.
Conceptually, the VDW-HRG model is quite different from an underlying fundamental QCD theory.
Still, the presented analysis is essentially parameter-free, 
in the sense that no new parameters which could be adjusted to lattice data were introduced.
Indeed, the two VDW parameters had been fixed by reproducing the saturation properties of nuclear matter~\cite{Vovchenko:2015vxa,Redlich:2016dpb}, independently from any lattice data. 
While there are other model parameters, e.g., the hadron list and its properties, they are known and fixed experimentally.
It is feasible that VDW parameters are different for different baryon pairs.
In the course of calculations, it was noticed that the agreement of the VDW-HRG with the lattice data is improved by taking smaller values of the nucleon or baryon EV parameter, $b \simeq 2-3$~fm$^3$.
Such modification does not necessarily break down the existing agreement of our model with the properties of nuclear matter: As suggested in Ref.~\cite{Alba:2016hwx}, the heavier and/or strange baryons may have smaller eigenvolumes, thus reducing the average $b$.
%Alternative ways of generalizing the attractive VDW interactions between nucleons to the full HRG can also be considered.
The present VDW-HRG model
leaves plenty of room for improvement.
Owing to the expectation that possible meson-related VDW interactions are considerably weaker than the baryon-baryon VDW interactions, we do not expect  major qualitative changes to the results presented in this paper for baryon number susceptibilities. This, however, may not be the case for other observables and should be carefully explored in future works.
%However, major qualitative changes to the results presented in this paper are not expected.

To summarize,
a minimal extension of the IHRG model
is presented which
includes both attractive and repulsive VDW interactions between baryons, with parameters $a$ and $b$ taken from previous fits to the ground state of nuclear matter.
%Parameters $a$ and $b$ have been taken from previous fits to the properties of the ground state of nuclear matter.
%Hence, the present VDW-HRG model is essentially parameter-free.
Compared to the usually used IHRG model,
the VDW-HRG model shows a qualitatively different behavior of most fluctuations and correlations of conserved charges in the ``crossover region'' at zero chemical potential.
This behavior resembles closely the lattice QCD results.
%The effect appears to be driven by both, repulsive interactions between (anti)baryons, and by the inclusion of VDW attraction which leads to a much improved agreement with lattice data.
These results hint towards crucial importance of the VDW interactions in the hadron gas, 
and
indicate that commonly performed comparisons of IHRG with the lattice data may result in misleading conclusions. 
Particularly, our results suggest that hadrons do not melt quickly with increasing temperature, as one could conclude on the basis of the IHRG.
It is feasible that the nuclear liquid-gas phase transition manifests itself into significant non-trivial fluctuations of net-baryon number in heavy-ion collisions. 
The influence of VDW interactions on thermal fits to hadron yield data from heavy-ion collisions is another possibility which will be explored.

%TC:ignore 

\begin{acknowledgments}

%\section*{Acknowledgements}
\emph{Acknowledgments.} We are grateful to Volker Koch for stimulating discussions, and
acknowledge fruitful comments from
Paolo Alba,
Leonid Satarov, and
Jan Steinheimer.
We also thank
Szabolcs Borsanyi for providing the preliminary data of Wuppertal-Budapest collaboration for the baryon-electric charge correlator in tabulated format.
This work was supported by HIC for FAIR within the LOEWE program of the State of Hesse.
V.V. acknowledges the support from HGS-HIRe for FAIR.
H.St. acknowledges the support through the Judah M. Eisenberg Laureatus Chair at Goethe University.
The work of M.I.G. was supported
by the Program of Fundamental Research of the Department of
Physics and Astronomy of National Academy of Sciences of Ukraine.

\end{acknowledgments}

%TC:endignore 


\begin{thebibliography}{999}

%\cite{Borsanyi:2013bia}
\bibitem{Borsanyi:2013bia}
  S.~Borsanyi, Z.~Fodor, C.~Hoelbling, S.~D.~Katz, S.~Krieg, and K.~K.~Szabo,
  %``Full result for the QCD equation of state with 2+1 flavors,''
  Phys.\ Lett.\ B {\bf 730}, 99 (2014).


%\cite{Bazavov:2014pvz}
\bibitem{Bazavov:2014pvz}
  A.~Bazavov {\it et al.} [HotQCD Collaboration],
  %``Equation of state in ( 2+1 )-flavor QCD,''
  Phys.\ Rev.\ D {\bf 90}, 094503 (2014).

\bibitem{Aoki:2006we}
  Y.~Aoki, G.~Endrodi, Z.~Fodor, S.~D.~Katz, and K.~K.~Szabo,
  %``The Order of the quantum chromodynamics transition predicted by the standard model of particle physics,''
  Nature {\bf 443}, 675 (2006).

%\cite{Borsanyi:2011sw}
\bibitem{Borsanyi:2011sw}
  S.~Borsanyi, Z.~Fodor, S.~D.~Katz, S.~Krieg, C.~Ratti, and K.~Szabo,
  %``Fluctuations of conserved charges at finite temperature from lattice QCD,''
  JHEP {\bf 1201}, 138 (2012).


%\cite{Bazavov:2012jq}
\bibitem{Bazavov:2012jq}
  A.~Bazavov {\it et al.} [HotQCD Collaboration],
  %``Fluctuations and Correlations of net baryon number, electric charge, and strangeness: A comparison of lattice QCD results with the hadron resonance gas model,''
  Phys.\ Rev.\ D {\bf 86}, 034509 (2012).


%\cite{Bellwied:2015lba}
\bibitem{Bellwied:2015lba}
  R.~Bellwied, S.~Borsanyi, Z.~Fodor, S.~D.~Katz, A.~Pasztor, C.~Ratti, and K.~K.~Szabo,
  %``Fluctuations and correlations in high temperature QCD,''
  Phys.\ Rev.\ D {\bf 92}, 114505 (2015).


%\cite{Bellwied:2013cta}
\bibitem{Bellwied:2013cta}
  R.~Bellwied, S.~Borsanyi, Z.~Fodor, S.~D.~Katz, and C.~Ratti,
  %``Is there a flavor hierarchy in the deconfinement transition of QCD?,''
  Phys.\ Rev.\ Lett.\  {\bf 111}, 202302 (2013).

\bibitem{Bazavov:2013dta} 
  A.~Bazavov {\it et al.},
  %``Strangeness at high temperatures: from hadrons to quarks,''
  Phys.\ Rev.\ Lett.\  {\bf 111}, 082301 (2013).

%\cite{Karsch:2013naa}
\bibitem{Karsch:2013naa}
  F.~Karsch,
  %``Thermodynamics of strong interaction matter from lattice QCD and the hadron resonance gas model,''
  Acta Phys.\ Polon.\ Supp.\  {\bf 7}, 117 (2014).

\bibitem{Borsanyi:2014ewa} 
  S.~Borsanyi, Z.~Fodor, S.~D.~Katz, S.~Krieg, C.~Ratti, and K.~K.~Szabo,
  %``Freeze-out parameters from electric charge and baryon number fluctuations: is there consistency?,''
  Phys.\ Rev.\ Lett.\  {\bf 113}, 052301 (2014).

\bibitem{Bazavov:2012vg} 
  A.~Bazavov {\it et al.},
  %``Freeze-out Conditions in Heavy Ion Collisions from QCD Thermodynamics,''
  Phys.\ Rev.\ Lett.\  {\bf 109}, 192302 (2012).

\bibitem{Stephanov:1998dy} 
  M.~A.~Stephanov, K.~Rajagopal, and E.~V.~Shuryak,
  %``Signatures of the tricritical point in QCD,''
  Phys.\ Rev.\ Lett.\  {\bf 81}, 4816 (1998).

%\cite{Fukushima:2014lfa}
\bibitem{Fukushima:2014lfa}
  K.~Fukushima,
  %``Hadron resonance gas and mean-field nuclear matter for baryon number fluctuations,''
  Phys.\ Rev.\ C {\bf 91}, 044910 (2015).


%\cite{Rischke:1991ke}
\bibitem{Rischke:1991ke}
  D.~H.~Rischke, M.~I.~Gorenstein, H.~Stoecker, and W.~Greiner,
  %``Excluded volume effect for the nuclear matter equation of state,''
  Z.\ Phys.\ C {\bf 51}, 485 (1991).


%\cite{BraunMunzinger:1999qy}
\bibitem{BraunMunzinger:1999qy}
  P.~Braun-Munzinger, I.~Heppe, and J.~Stachel,
  %``Chemical equilibration in Pb + Pb collisions at the SPS,''
  Phys.\ Lett.\ B {\bf 465}, 15 (1999).


%\cite{Vovchenko:2015xja}
\bibitem{Vovchenko:2015xja}
  V.~Vovchenko, D.~V.~Anchishkin, and M.~I.~Gorenstein,
  %``Particle number fluctuations for the van der Waals equation of state,''
  J.\ Phys.\ A {\bf 48}, 305001 (2015).


%\cite{Vovchenko:2015vxa}
\bibitem{Vovchenko:2015vxa}
  V.~Vovchenko, D.~V.~Anchishkin, and M.~I.~Gorenstein,
  %``Van der Waals Equation of State with Fermi Statistics for Nuclear Matter,''
  Phys.\ Rev.\ C {\bf 91}, 064314 (2015).


%\cite{Vovchenko:2015pya}
\bibitem{Vovchenko:2015pya}
  V.~Vovchenko, D.~V.~Anchishkin, M.~I.~Gorenstein, and R.~V.~Poberezhnyuk,
  %``Scaled variance, skewness, and kurtosis near the critical point of nuclear matter,''
  Phys.\ Rev.\ C {\bf 92}, 054901 (2015).


%\cite{Redlich:2016dpb}
\bibitem{Redlich:2016dpb}
  K.~Redlich and K.~Zalewski,
  %``Thermodynamics of Van der Waals Fluids with quantum statistics,''
  Acta Phys. Polon. B \textbf{47}, 1943 (2016),
  [arXiv:1605.09686 [cond-mat.quant-gas]].

%\cite{Andronic:2012ut}
\bibitem{Andronic:2012ut}
  A.~Andronic, P.~Braun-Munzinger, J.~Stachel, and M.~Winn,
  %``Interacting hadron resonance gas meets lattice QCD,''
  Phys.\ Lett.\ B {\bf 718}, 80 (2012).
  
%\cite{Vovchenko:2014pka}
\bibitem{Vovchenko:2014pka}
  V.~Vovchenko, D.~V.~Anchishkin, and M.~I.~Gorenstein,
  %``Hadron Resonance Gas Equation of State from Lattice QCD,''
  Phys.\ Rev.\ C {\bf 91}, 024905 (2015).

\bibitem{Venugopalan:1992hy} 
  R.~Venugopalan and M.~Prakash,
  %``Thermal properties of interacting hadrons,''
  Nucl.\ Phys.\ A {\bf 546}, 718 (1992).

%\cite{Agashe:2014kda}
\bibitem{Agashe:2014kda}
  K.~A.~Olive {\it et al.} [Particle Data Group Collaboration],
  %``Review of Particle Physics,''
  Chin.\ Phys.\ C {\bf 38}, 090001 (2014).


%\cite{Broniowski:2015oha}
\bibitem{Broniowski:2015oha}
  W.~Broniowski, F.~Giacosa, and V.~Begun,
  %``Cancellation of the $\sigma$ meson in thermal models,''
  Phys.\ Rev.\ C {\bf 92}, 034905 (2015).


%\cite{Friman:2015zua}
\bibitem{Friman:2015zua}
  B.~Friman, P.~M.~Lo, M.~Marczenko, K.~Redlich, and C.~Sasaki,
  %``Strangeness fluctuations from $K-\pi$ interactions,''
  Phys.\ Rev.\ D {\bf 92}, 074003 (2015).


%\cite{Vovchenko:2015cbk}
\bibitem{Vovchenko:2015cbk}
  V.~Vovchenko and H.~Stoecker,
  %``Heavy-ion hot hadron matter freezes out at T=170-320 MeV?,''
  %arXiv:1512.08046 [hep-ph],
  %J. Phys. G, in print.
  J.\ Phys.\ G {\bf 44}, 055103 (2017).


%\cite{Vovchenko:2015idt}
\bibitem{Vovchenko:2015idt}
  V.~Vovchenko, V.~V.~Begun, and M.~I.~Gorenstein,
  %``Hadron multiplicities and chemical freeze-out conditions in proton-proton and nucleus-nucleus collisions,''
  Phys.\ Rev.\ C {\bf 93}, 064906 (2016).





% %\cite{Vovchenko:2014pka}
% \bibitem{Vovchenko:2014pka}
%   V.~Vovchenko, D.~V.~Anchishkin and M.~I.~Gorenstein,
%   %``Hadron Resonance Gas Equation of State from Lattice QCD,''
%   Phys.\ Rev.\ C {\bf 91}, 024905 (2015).


%\cite{Lo:2015cca}
\bibitem{Lo:2015cca}
  P.~M.~Lo, M.~Marczenko, K.~Redlich, and C.~Sasaki,
  %``Matching the Hagedorn mass spectrum with Lattice QCD results,''
  Phys.\ Rev.\ C {\bf 92}, 055206 (2015).


%\cite{Bazavov:2014xya}
\bibitem{Bazavov:2014xya}
  A.~Bazavov {\it et al.},
  %``Additional Strange Hadrons from QCD Thermodynamics and Strangeness Freezeout in Heavy Ion Collisions,''
  Phys.\ Rev.\ Lett.\  {\bf 113}, 072001 (2014).

\bibitem{Borsanyi:2015axp} 
  S.~Bors\'anyi,
  %``Fluctuations at finite temperature and density,''
  PoS LATTICE {\bf 2015}, 015 (2016)
  [arXiv:1511.06541 [hep-lat]].
  
\bibitem{WBchi11BQ} 
  Wuppertal-Budapest collaboration, in preparation.

\bibitem{Alba:2014eba} 
  P.~Alba, W.~Alberico, R.~Bellwied, M.~Bluhm, V.~Mantovani Sarti, M.~Nahrgang, and C.~Ratti,
  %``Freeze-out conditions from net-proton and net-charge fluctuations at RHIC,''
  Phys.\ Lett.\ B {\bf 738}, 305 (2014).

\bibitem{Aggarwal:2010wy} 
  M.~M.~Aggarwal {\it et al.} [STAR Collaboration],
  %``Higher Moments of Net-proton Multiplicity Distributions at RHIC,''
  Phys.\ Rev.\ Lett.\  {\bf 105}, 022302 (2010).
  
\bibitem{Adamczyk:2013dal} 
  L.~Adamczyk {\it et al.} [STAR Collaboration],
  %``Energy Dependence of Moments of Net-proton Multiplicity Distributions at RHIC,''
  Phys.\ Rev.\ Lett.\  {\bf 112}, 032302 (2014).

\bibitem{Luo:2015ewa} 
  X.~Luo [STAR Collaboration],
  %``Energy Dependence of Moments of Net-Proton and Net-Charge Multiplicity Distributions at STAR,''
  PoS CPOD {\bf 2014}, 019 (2015)
  [arXiv:1503.02558 [nucl-ex]].

\bibitem{Mukherjee:2016nhb} 
  A.~Mukherjee, J.~Steinheimer, and S.~Schramm,
  %``Higher-order baryon number susceptibilities: interplay between the chiral and the nuclear liquid-gas transitions,''
  arXiv:1611.10144 [nucl-th].

\bibitem{Satarov:2016peb} 
  L.~M.~Satarov, V.~Vovchenko, P.~Alba, M.~I.~Gorenstein, and H.~Stoecker,
  %``New scenarios for hard-core interactions in a hadron resonance gas,''
  Phys.\ Rev.\ C {\bf 95}, 024902 (2017).

%\cite{Alba:2016hwx}
\bibitem{Alba:2016hwx}
  P.~Alba, V.~Vovchenko, M.~I.~Gorenstein, and H.~Stoecker,
  %``Flavor-dependent eigenvolume interactions in hadron resonance gas and its implications for hadron yields at LHC energies,''
  arXiv:1606.06542 [hep-ph].


\end{thebibliography}
\end{document}